# Ionic-strength and pH dependent reactivities of ascorbic acid toward ozone in aqueous micro-droplets studied by aerosol optical tweezers

Yuan-Pin Chang,[1,2,3*] Shan-Jung Wu,[1] Min-Sian Lin,[1] Che-Yu Chiang,[1] Genin Gary Huang[3]

1. Department of Chemistry, National Sun Yat-sen University, Kaohsiung 80424, Taiwan
2. Aerosol Science Research Center, National Sun Yat-sen University, Sizihwan, Kaohsiung 80424, Taiwan
3. Department of Medicinal and Applied Chemistry, Kaohsiung Medical University, Kaohsiung 80708, Taiwan

* E-mail: ypchang@mail.nsysu.edu.tw

## Abstract

The heterogeneous oxidation reaction of single aqueous ascorbic acid ($AH_2$) aerosol particles with gas-phase ozone was investigated in this study utilizing aerosol optical tweezers with Raman spectroscopy. The measured liquid-phase bimolecular rate coefficients of the $AH_2$ + $O_3$ reaction exhibit a significant pH dependence, and the corresponding values at ionic strength 0.2 M are $(3.1 \pm 2.0) \times 10^5$ $M^{-1}s^{-1}$ and $(1.2 \pm 0.6) \times 10^7$ $M^{-1}s^{-1}$ for pH $\approx$ 2 and 6, respectively. These results measured in micron-sized droplets agree with those from previous bulk measurements, indicating that the observed aerosol reaction kinetics can be solely explained by liquid phase diffusion and $AH_2$ + $O_3$ reaction. Furthermore, the results indicate that high ionic strengths could enhance the liquid-phase rate coefficients of the $AH_2$ + $O_3$ reaction. The results also exhibit a negative ozone pressure dependence that can be rationalized in terms of a Langmuir–Hinshelwood type mechanism for the heterogeneous oxidation of $AH_2$ aerosol particles by gas-phase ozone. The results of the present work imply that in acidified airway-lining fluids the antioxidant ability of $AH_2$ against atmospheric ozone will be significantly suppressed.

## Introduction

Ascorbic acid ($AH_2$) is one of main antioxidants in epithelial lining fluids (ELF) which work as thin fluid layer (typical ~0.1-0.2 μm thick) on the surfaces of airways and alveoli to protect against atmospheric oxidants, such as ozone. The reaction pathways of the ozonolysis of aqueous $AH_2$ (pKa = 4.1) significantly depend on pH. For the case of pH > pKa, such as pH ≈ 7, the dominant form of $AH_2$ is monoanion, $AH^-$, and the ozonolysis of $AH^-$ produces dehydroascorbic acid (DHA) and singlet $O_2$ in high yields (> 90%):[1,2]

$$AH^- + O_3 \rightarrow DHA + O_2(^1\Delta_g) + OH^-  \qquad (1)$$

The singlet $O_2$ product as secondary oxidant may transduce oxidation damages through ELF,[3] as it is well known to react actively with biomolecules in proteins and DNA.[4,5] On the other hand, the ozonolysis of $AH_2$ in the free acid form firstly produces an unstable primary 1,2,3-trioxolane ozonide (POZ), which could subsequently subject to unimolecular decomposition and form a secondary ascorbate ozonide (AOZ) or threonic acid (THR):[1]

$$AH_2 + O_3 \rightarrow POZ \rightarrow AOZ \text{ or } THR  \qquad (2)$$

AOZ may also be qualified as harmful secondary oxidant, as the finding of its short lifetime in bulk water implies its potential high reactivity.[1]

Besides the reaction mechanisms, pH can also affect the reactivities of $AH_2$ toward ozone. The reaction rate coefficients of the $AH_2$ + ozone reaction at different pH have been determined by various experimental studies, while the measurement results seem to highly depend on the experimental designs or conditions. The measurements in bulk water carried out by Giamalva et al.[6] show that the rate coefficients at pH = 2.0 and 4.8 are $6.9 \times 10^5$ and $5.6 \times 10^7$ $M^{-1}s^{-1}$, respectively. The measurements of Kanofsky and Sima[7] utilizing a type of reactor where the liquid contacted with gaseous ozone yielded similar values: at pH = 2.0 and 7.0 the measured rate coefficients were $5.6 \times 10^5$ and $4.8 \times 10^7$ $M^{-1}s^{-1}$, respectively. Two later studies[8,9] utilized the similar type of reactors but carried out the reactions in ELF model solutions (pH = 7.4) instead. However, the measured rate coefficients were significantly smaller than above reported values of pH > pKa: $5.5 \times 10^4$ M$^{-1}$s$^{-1}$ from Kermani et al.[9] Recently, Enami et al.[1] studied this reaction in micro-droplets as a model system of air-water interfaces. The aqueous AH$_2$ droplets in size of few μm were exposed to gaseous ozone for very short time, and then the composition of the interfacial layers of reacting droplets was analyzed by online electrospray mass spectrometry. They observed that the reaction rates at the interfacial layer (a few nanometers) were at least two orders of magnitude larger than those from bulk measurements of Giamalva et al. On the other hand, their observed ratio of reaction rate coefficients at pH > pKa and pH < pKa was about one order of magnitude smaller than that in bulk water.[1,6]

In this study, we utilized an aerosol optical tweezers (AOT) apparatus, which can trap single liquid droplets via optical gradient force, to study the reactive uptake processes of single micro-droplets of aqueous AH$_2$ to gaseous ozone. The interaction of micro-droplets with gaseous environments can be a model system for the interaction between gas and fluid films[1] and exhaled bioaerosols, as the reactive uptake behavior of micro-droplets involves the dynamics processes of both the bulk phase and the gas-liquid interface.[10] Also the rate equations and kinetics models for interpreting the reactive uptakes of aerosol droplets have been well established, providing a straightforward strategy to retrieving the rate coefficients of interest from experimental data of AOT directly.[10,11] Particularly, one of the main advantages of AOT is that the microphysical properties of each single trapped micro-droplet, such as radius, refractive index, viscosity, diffusion coefficient, surface tension, temperature, pH, vapor pressure and compositions, can be determined via spectroscopic means in high accuracy and in real time.[12–25] Thus, AOT can be utilized as a wall-less reactor to investigating heterogeneous reactions in details, as the reaction kinetics and microphysical properties of each particle can be monitored simultaneously.[26]

There have been several experimental investigations of the reaction kinetics of heterogeneous oxidations in single droplets via AOT combined with Raman spectroscopy.[26–31] King et al.[27] studied the ozonolysis of aqueous and organic droplets containing unsaturated organic compounds and obtained their reactive uptake coefficients, and the liquid-phase bimolecular rate coefficients retrieved from the reactive uptake coefficients agreed with literature values. Dennis-Smither et al.[28] studied the heterogeneous oxidation of aqueous maleic acid droplets by ozone, and their measured reactive uptake coefficients agreed with literature values. Recently, Hunt et al.[26] investigated the heterogeneous oxidation of nitrite anion in aqueous droplets by ozone, attempting to clarify whether the surface excess or depletion of nitrite anions could affect the aerosol reaction kinetics. Their measured rate coefficient agreed with literature values of bulk measurements, implying that it is not necessary to include the surface effects in the reaction kinetics of aqueous nitrite droplets under atmospheric conditions.

In this work we utilized AOT combined with Raman spectroscopy to study the kinetics of the reaction of aqueous $AH_2$ micro-droplets with gaseous ozone. We measured the bimolecular reaction rate coefficients at different values of pH, ozone pressures and ionic strengths, and their effects to the rate coefficients of the $AH_2$ + ozone reaction will be addressed in this report. Finally, we discuss the implications of these results for antioxidant kinetics in ELF and exhaled bioaerosols.

## Experimental

The aerosol optical tweezers apparatus used in this work is similar to those of previous studies of AOT,[32] consisting a trapping laser, an inverted microscope and a Raman spectrometer. We utilized a CW Nd:YVO$_4$ laser operating at 532 nm (Coherent, Verdi V2) as the trapping laser. The typical laser power before entering the microscope was about 30 to 130 mW. For achieving an optimal trapping force, the laser beam was expanded by two plano-convex lenses. The expanded laser beam was guided into the epi-illumination port of the inverted microscope (Nikon, Eclipse Ti2), and then it was reflected to an oil-immersion objective (Nikon, CFI Plan Apochromat lambda, 60X, NA of 1.4, WD of 0.13 mm) by a dichroic mirror (Chroma, ZT532dcrb). The focused light passed from the objective through an index matching fluid and through a coverslip into an aerosol trapping chamber that was mounted on the sample stage of the microscope. The Raman scattering light emitted from the trapped droplet was collected by the objective, passing through a holographic notch filter (Semrock, NF03-532E-25), and imaged onto the entrance slit of a 0.5 m spectrograph (Andor, SR-500i, using 1200 l/mm grating in this work) coupled with a TE-cooled spectroscopic CCD (Andor, DU401-BVF). The integrated time for acquiring Raman spectra was 1 second. The brightfield image of the trapped droplet was acquired by utilizing a 455 nm LED (Thorlabs, M455L3) as the light source, a CCD camera (Basler, ICDA-ACA640-90UM), and a 451 nm bandpass filter (Semrock, FF01-451/106-25).

A dense flow of aqueous L-ascorbic acid (Sigma-Aldrich) aerosol was generated by a medical nebulizer (Sumo, V-15), and the sizes of aerosol droplets were about 3 to 8 μm in diameter. The aerosol flow was introduced into the trapping chamber, and after waiting about a few minutes, a single droplet will be caught by the focused laser. The relative humidity (RH) in the trapping chamber was maintained around 85% *via* injecting a wet nitrogen gas flow with a flow rate about 300 sccm, which was controlled by a mass flow controller (MKS, 1179A), and it was monitored by a humidity sensor (Honeywell, HIH-4602-C, accuracy: 3.5%) which was calibrated by a more accurate one (Rotronic, HC2A-S, accuracy: 0.5%). To maintain this high RH inside the trapping chamber, all of its components were immersed in water prior experiments. After the chamber was assembled, it was further flashed with the wet nitrogen flow at least over 30 minutes, before starting the trapping experiments. During a typical experimental time, such as a couple of hours, ~85% RH in the chamber may slowly drift around 5%, while supplying the constant wet nitrogen flow. We chose not to change the flow rate to regulate RH during trapping experiments, as a stable trapping is also affected by the gas flow. The temperature inside the chamber was maintained at 297 K. An ozone generator (Airphysics, C-L010-DSI) was used to produce gaseous $O_3$ from a pure $O_2$ gas flow, and the generated $O_2/O_3$ gas mixture was further diluted with dry $N_2$ in a stainless steel chamber. During the reaction kinetics measurements, the premixed $O_2/O_3/N_2$ gas flow of few sccms was mixed with the wet $N_2$ gas flow prior flowing into the trapping chamber. The concentration of ozone was on-line monitored via its UV absorption peak at 250 nm by using an absorption cell (length of 75 cm), a deuterium lamp (Hamamatsu, S2D2 module), and a CCD spectrometer (Ocean Optics, Flame-S-UV-VIS-ES). For the measurements at pH ≈ 2, the solution for generating aerosol contained 0.3 M $AH_2$ and 0.1 M NaCl. For the case of pH ≈ 6, the 0.3 M $AH_2$ solution contained 0.5 M sodium phosphate buffer and 0.1 M NaCl, and pH of the final solution was adjusted to 6.0 using solutions of HCl and NaOH.

## Results

### Raman spectra of trapped droplets

Figure 1 shows the representative Raman spectra of a trapped aqueous $AH_2$ droplet in pH ≈ 2, before and after the reaction with ozone. The measured Raman bands of aqueous $AH_2$, such as the coupled C=C and C=O stretching mode at 1690 $cm^{-1}$, indicated that the dominate form of aqueous $AH_2$ is free acid.[33] After the trapped $AH_2$ droplets were exposed to ozone, the Raman peak at 1690 $cm^{-1}$ gradually decreased, due to the cycloaddition of ozone toward the C=C double bond of $AH_2$.[1,34] The relatively broad bands of products peaked at around 1790 $cm^{-1}$ and 3000 $cm^{-1}$ were also formed, as shown in Figure 1, and they were tentatively assigned to C=O and C-H stretching modes of threonic acid, respectively. Enami at al. observed that the other reaction product AOZ depicted in reaction (2) only survived on the air-water interface due to its short lifetime in water.[1] These interfacial AOZ were expected to have relatively low concentration and thus not observed in this study because of the limited sensitivity of Raman spectroscopy. For the trapped $AH_2$ droplets with pH ≈ 6, the coupled C=C and C=O stretching mode of $AH^-$ is centered at 1590 $cm^{-1}$, as shown in Figure S1.[33] After the $AH^-$ droplets were exposed to ozone, several Raman bands of products were formed, such as those centered at 1400 $cm^{-1}$, 1720 $cm^{-1}$ and 3000 $cm^{-1}$ (see Figure S1 and S2), and they were tentatively assigned to dehydroascorbic acid, which is the main product of $O_3$ + $AH^-$ reaction depicted in reaction (1).

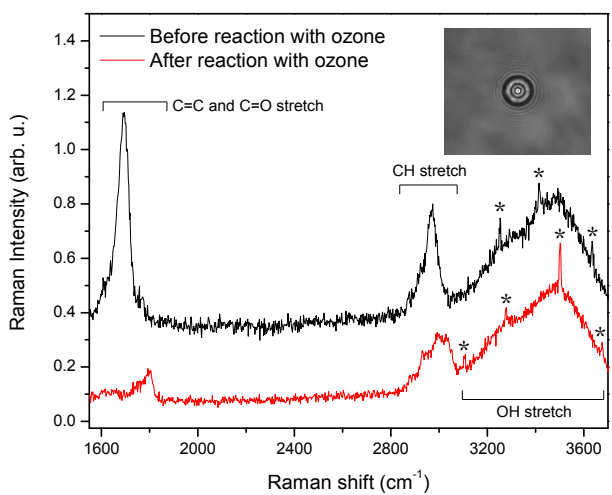

**Figure 1.** Representative Raman spectra of an optically trapped aqueous $AH_2$ droplet in pH = 1.8 before (top data) and after (bottom data) the reaction with ozone at 85% RH. Two spectra are offset for clarity. Note that relatively sharp peaks in the spectra are CERS signals (labelled with asterisk). The inset shows the brightfield image of the droplet before the reaction with ozone. See Table S1 (ESI) (expt. 01) for details of the experimental conditions.

## Determining microphysical properties of trapped droplets

In the Raman spectra of aqueous $AH_2$ droplets there are also several relatively sharp peaks superposed on the relatively broad molecular Raman band (see Figures 1 and S1). They are so called cavity-enhanced Raman scattering (CERS), which are whisper-gallery modes amplified by stimulated Raman scattering. [32,35,36] These CERS signals can be used to retrieve the values of radius and refractive index (RI) of the droplet with a high accuracy. We utilized a Mie theory simulation program *mrfit* developed by Preston et al.[13] to analyse the spectral positions of CERS signals and obtained fitted RI and radius. The fitted RI values were further used to retrieve the absolute concentration of $AH_2$ before the reaction, $[AH_2]_0$, and ionic strength inside the droplet. We also utilized the time evolution of fitted radius to track the change of droplet ionic strength during a reaction progress. When no obvious CERS signals existed, the radius and $[AH_2]_0$ of the trapped droplet were determined by brightfield imaging and Raman intensity ratios of $AH_2$ and water, respectively. As all solutes in droplets, such as $AH_2$, $Na^+$, $Cl^-$ and phosphates, are involatile, the droplet ionic strength can also be determined *via* comparing the concentrations of $AH_2$ in droplets and in the bulk solution. The values of $[AH_2]_0$ determined by CERS and spontaneous Raman spectroscopy are also in good agreement, justifying both methods utilized in this study.

Tables S1 and S2 list the values of radius, $[AH_2]_0$ and droplet ionic strength determined by above methods for each trapped droplet. The typical droplet radius in the present work is about 3 μm. Both $[AH_2]_0$ and ionic strength of trapped droplets were found to be several times larger than those of the bulk solution on average, probably caused by a vaporizing enrichment before flowing into the moisturized trapping chamber.

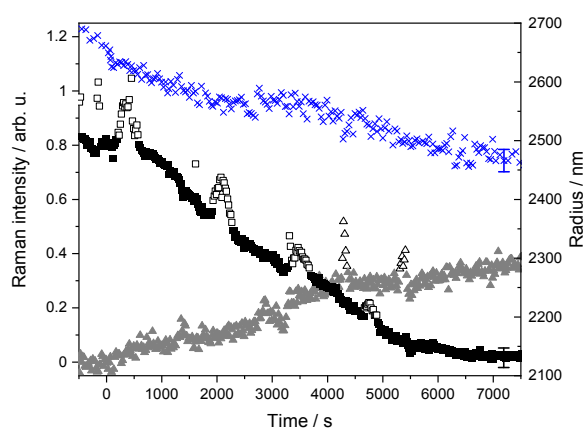

**Figure 2.** Representative time profiles of radius (crosses) and integrated Raman intensities of a trapped $AH_2(aq)$ droplet which was exposed to gaseous ozone after 0th second. The Raman intensities of $AH_2$ (squares) and products (triangles) were obtained from integrating the areas of the Raman peaks centered at 1690 cm$^{-1}$ and 1790 cm$^{-1}$, respectively. Most intensity jumps in the Raman intensity time profiles are ascribed to drifting CERS peaks. Hollow symbols represent the intensity interferences caused by frequency-drifting CERS peaks. For clarity, only the typical error bars are provided at the end of data. The ozone pressure is 1.13 ppm. See Table S1 (ESI) (expt. 17) for details of the experimental conditions.

## Determining pH of trapped droplets

For the kinetics measurements at pH ≈ 6, we estimated the values of pH inside the trapped droplets via comparing molecular Raman signals of inorganic acids and their conjugate bases, following the methods developed by previous works utilizing sodium bisulfate buffer. We utilized the Raman peak area ratios of $H_2PO_4^-$ (874 cm$^{-1}$) and $HPO_4^{2-}$ (987 cm$^{-1}$) (see Figure S3) to determine pH over the range of 5.5 to 7.3, with the calibration curve shown in Figure S4. Note that the Raman intensities of $H_2PO_4^-$ and $HPO_4^{2-}$ at pH ≈ 6 are very similar, facilitating the determination of pH around this range. The pH measurement results show that the values of pH inside droplets before exposed to ozone were equal to about 6.2 on average. After the reactions with ozone finished, the pH values of droplets slightly increased to about 6.7 on average (see Table S2), agreeing with the predicted mechanism of reaction (1). For the case of pH ≈ 2, the values of pH before reaction were estimated from the measured values of [AH$_2$] and known pKa (see Table S1).

It should be noted that the above estimations of pH performed by the present work did not consider the effect of high ionic strengths inside micro-droplets. Two recent studies utilizing AOT to determine aerosol pH have demonstrated that the non-ideal thermodynamics due to high ionic strengths has to be taken into account *via* thermodynamic model calculations, such as specific ion interaction theory or E-AIM aerosol thermodynamics model, in order to predict the proton activity correctly.[23,25] Otherwise, the estimated values of pH can be less accurate. For the case of aqueous sulfate/bisulfate systems, such deviation could be up to about 2 pH units.[23,25] We assume that this may be regarded as the maximum deviation of pH reported in this work. The main limitation of the present work to predict the proton activity more appropriately is a lack of the thermodynamic coefficients of AH$_2$ required by the non-ideal thermodynamic models described above, and lots of experimental and theoretical efforts are still needed for such purpose.

### Kinetics of the AH$_2$ + ozone reaction

Figures 2 and S5 show the time evolutions of radius and Raman intensities of the aqueous AH$_2$ droplets exposed to ozone at pH ≈ 2 and pH ≈ 6, respectively. After exposed to ozone at 0 second, the Raman intensities of AH$_2$ decreased immediately, and the product Raman bands emerged at the same time. Their Raman intensity time profiles further show that the formation of products was concomitantly accompanied with the decay of AH$_2$, indicating that the observed product was solely created from the AH$_2$ + ozone reaction. This observation also indicates that the products are inert toward ozone, agreeing with the findings of Enami et al.[1] The Raman intensity time profiles in Figure 2 also exhibit several intensity fluctuations for short periods of time, and they are caused by the progress of CERS peaks through the corresponding molecular Raman bands.[28] In data analysis, such CERS interference was identified by tracing the progress of CERS peaks in Raman spectra time series and labelling the times when a CERS peak superposes with the molecular Raman band of interest, such as the data points labelled with hollow symbols in Figure 2. Such assignment of progressing CERS peaks can be assisted and justified by Mie theory simulation, when assuming continuous changes of droplet size and RI and no mode hopping of CERS peaks. Figure 2 also shows that the radius only decreased slightly during the reaction progress, implying involatile reaction products and slow evaporation of water. For simplifying the following kinetics analysis, the radius is assumed to be a fixed value in the data analysis, and we assume the change of radius during the reaction to be the error of radius.

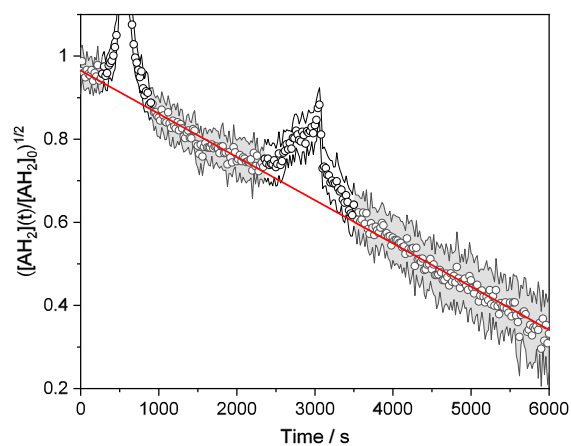

**Figure 3.** Square root of the normalized Raman intensities of the $AH_2$ C=C band versus reaction time (symbols) at pH = 1.8 and ozone pressure $P$ = 1.4 ppm (expt. 19) and the fit (straight line) based on equation (3). Each data point represents the average over 20 consecutive spectra. Two rugged lines represent an error band which corresponds to 1σ uncertainties of data. The significant bursts of intensities (symbols not in grey areas) were attributed to frequency-drifting CERS peaks, and thus they were not included in the fit.

According to Smith et al.[10] and King et al.,[27] the reactive uptake of gaseous ozone on the aqueous $AH_2$ droplet studied in the present work can be approximated as the diffusion-limited special case, where the dissolved aqueous ozone reacts with the aqueous $AH_2$ near the surface and thus the reaction rate depends on the bimolecular rate coefficient of the $AH_2$ + ozone reaction in aqueous phase ($k_2$) and the diffusion coefficient of aqueous ozone ($D_0$). The integrated rate equation can be expressed as following:

$$\sqrt{\frac{[AH_2](t)}{[AH_2]_0}} = 1 - \frac{3HP}{2r}\sqrt{\frac{D_0 k_2}{[AH_2]_0}}\, t \qquad (3)$$

where $[AH_2](t)$ is $[AH_2]$ as function of $t$, $[AH_2]_0$ represents $[AH_2]$ at 0 second of reaction time, $H$ is the Henry's Law constant of ozone in the aqueous solution, $P$ is the pressure of ozone and $r$ is the radius of the droplet. The value of $D_0$ is about $1.8 \times 10^{-9}$ m$^2$ s$^{-1}$.[35] The values of $H$ at various solute concentrations used here can be estimated from a Sechenov relation (see ESI).[38] According to equation (3), a plot of $([AH_2]/[AH_2]_0)^{1/2}$ versus $t$ should yield a straight line, as shown in Figure 3, where $[AH_2]$ is represented by the C=C stretching mode Raman intensity of $AH_2$, and the fitted gradient can be used to derive the value of $k_2$. The values of $k_2$ determined from all data of this study are summarized in Tables S1 and S2, for pH ≈ 2 and ≈ 6, respectively. Finally, note that we only fit the data within the reaction time where the values of $([AH_2](t)/[AH_2]_0)^{1/2}$ are between 1 and ~0.3 (see Figure 3), because of significant standard deviations of data at the longer reaction times. Note that the CERS interferences described previously can cause significant deviations of the Raman intensity time profile from Equation (3), as shown in Figure 3. Thus, the data points used to derive $k_2$ did not include those associated with CERS interferences, such as the symbols not in grey areas in Figure 3. Finally, for the case of pH ≈ 2 measurements, when assuming that pH of the trapped droplet was solely regulated by $[AH_2]$, the estimated change of aerosol pH in this fitting range is about 0.5, which defines the uncertainty of pH for the fitted values of $k_2$.

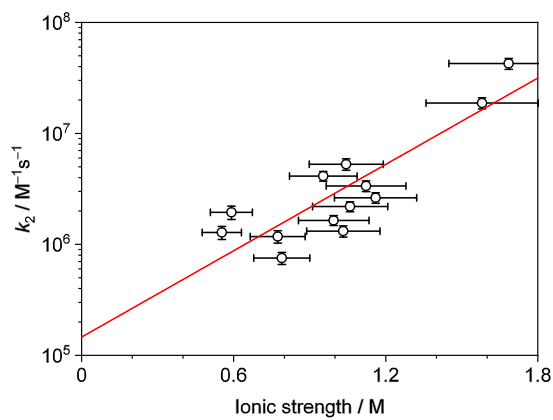

**Figure 4.** Bimolecular reaction rate coefficient $k_2$ (symbols) of pH ≈ 2 plotted as a function of droplet ionic strength and the fit (line) using equation (4). The pressure of ozone applied to each data is below 3.5 ppm. Each plotted error bar includes all experimental errors (see ESI).

## Discussion

### Effect of ionic strength on bimolecular rate coefficients

To clarify the effects of ionic strength, we measured the reaction rate coefficient $k_2$ at different ionic strengths. Figure 4 plots all the measured rate coefficients of pH ≈ 2 as a function of ionic strength, indicating a positive correlation between ionic strengths and reaction rate coefficients. When the ionic strengths increase from 0.6 M to 1.6 M, which are contributed by enriched concentrations of NaCl in the trapped droplets, the reaction rate coefficients can increase about ten times. According to Laidler, the rate coefficient $k$ of a reaction between an ion and an neutral molecule, <span style="color:red">or between two neutral molecules,</span> at ionic strength $I$ may have an approximate relation with the rate coefficient at zero ionic strength $k_0$:[39]

$$\log_{10}k = log_{10}k_0 + b'I \qquad (4)$$

where $b'$ is an empirical constant. Figure 4 also shows the fitting based on equation (4) to data of pH ≈ 2 at droplet ionic strength 0.5 M to 1.7 M, which was dominated by enriched NaCl, and the fitted values of $k_0$ and $b'$ are $(1.7 ± 1.1) \times 10^5$ M$^{-1}$s$^{-1}$ and 1.2 ± 0.2 M$^{-1}$, respectively. Figure 5 plots all the measured rate coefficients of pH ≈ 6 at droplet ionic strength 1.7 M to 5.3 M, which were mostly contributed by enriched AH$^-$, Na$^+$ and phosphates in the trapped droplets. Similar to data of pH ≈ 2, the reaction rate coefficients of pH ≈ 6 also exhibit a weak positive dependence with ionic strengths. We also applied equation (4) to fit these data of pH ≈ 6, as shown in Figure 5. The fitted values of $k_0$ and $b'$ are $(1.1 \pm 0.6) \times 10^7$ M⁻¹s⁻¹ and $0.23 \pm 0.04$ M⁻¹, respectively.

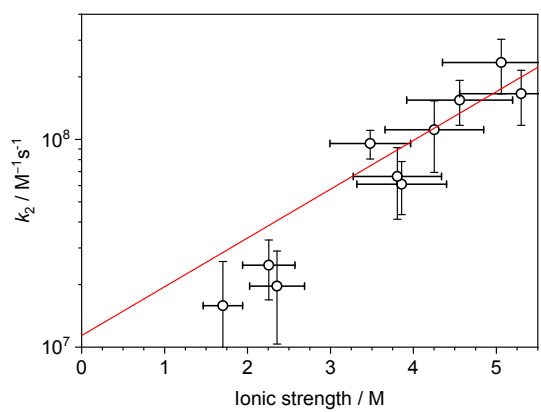

**Figure 5.** Bimolecular reaction rate coefficient $k_2$ (symbols) of pH ≈ 6 plotted as a function of droplet ionic strength and the fit (line) using equation (4). The pressure of ozone applied to each data is below 1.5 ppm. Each plotted error bar includes all experimental errors (see the Supporting Information).

The activity coefficients of solutes in liquid phase can be affected by the ionic strength, and such effect for reactions containing neutral molecules is considered in Equation (4) by means of Debye-McAulay approach.[39] Base on such framework, the positive ionic-strength dependence could be ascribed to increases in activity coefficients of reactants or/and the stabilization of the activated complex which are approximated by the $b'I$ term in Equation (4).[40,41] The reaction of ozone and $AH_2$ involves an attack of ozone to the carbon-carbon double bond of $AH_2$, the corresponding pre-reactive complex could have a larger dipole moment than reactants. Thus, under conditions of high ionic strengths, the activated complex could be more stabilized than reactants due to its larger ion-dipole interactions.[41] Such effect can further result in a lower energy barrier and thus a faster reaction rate. For both cases of pH ≈ 2 and 6, we assume that this effect may play a role in their positive ionic-strength dependences. However, for the case of pH ≈ 2, the increase in the ionic strength could also increase the activity coefficient of $AH_2$ in the free acid form and thus decrease its solubility, so called salting-out effect.[42] When increasing the ionic strength, the water molecules in the solvation shell of $AH_2$ will be displaced by salt ions, reducing the available volume of the aqueous solution to dissolve $AH_2$. In aerosol phase, such effect could further cause $AH_2$ molecules to be repelled to the water-air interface, and the concentration of $AH_2$ near the surface would be higher than in bulk phase, enhancing the reactions of $AH_2$ with ozone near the surface. Such surface enrichment due to increasing the ionic strength has been investigated by several previous studies.[40,43–45] For the case of pH ≈ 6, this salting-out effect probably does not play a role, as the increase of the ionic strength can decrease the activity coefficient and thus increase the solubility of $AH^-$ as charged species instead. This may partially explain why the fitted value of $b'$ for pH ≈ 2 is larger than that for pH ≈ 6.

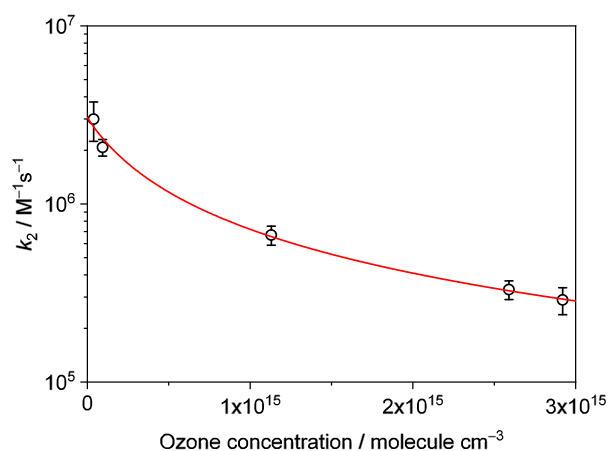

**Figure 6.** Bimolecular reaction rate coefficients $k_2$ (symbols) of pH ≈ 2 and droplet ionic strength ≈ 1 M as a function of gas-phase ozone concentration and the fit (line) utilizing Equation (5). The plotted error bar of each rate coefficient corresponds to 1σ statistic errors.

**Effect of gaseous ozone pressure on bimolecular rate coefficients**

We also measured the reaction rate coefficients $k_2$ at different gas-phase ozone pressures. Figure 6 plots the rate coefficients of the same ionic strength at pH ≈ 2 and from $P$ ≈ 2 ppm to 100 ppm. The higher ozone pressures yield the smaller values of $k_2$. Such ozone pressure dependence may be attributed to the surface saturation effects for reactants, so called Langmuir-Hinshelwood mechanism, which could cause the decrease of reactive uptake coefficients and the underestimation of the measured $k_2$.[46,47] The relationship between the measured rate coefficient $k_2$ and the gas-phase ozone pressure $P$ for the Langmuir-Hinshelwood-type mechanism can be modelled using:[39]

$$k_2 = \frac{k_{2,\max}}{1 + K_{\text{ozone}}P} \qquad (5)$$

Where $K_{\text{ozone}}$ is the ozone gas-to-surface equilibrium constant and $k_{2,\max}$ is the maximum bimolecular rate coefficient measured at the low ozone pressure limit. Figure 6 shows that the fit based on Equation (5) has a good agreement with the data. The fitted values for $k_{2,\max}$ and $K_{\text{ozone}}$ are $(3.1 \pm 0.3) \times 10^6$ M$^{-1}$s$^{-1}$ and $(3.2 \pm 0.3) \times 10^{-15}$ cm$^3$molecule$^{-1}$, respectively. The fitted value of $K_{\text{ozone}}$ for aqueous AH$_2$ droplets is similar to those previously reported on aqueous aerosol particles of maleic acid ($(9 \pm 4) \times 10^{-15}$ cm$^3$molecule$^{-1}$) and fumaric acid ($(5 \pm 2) \times 10^{-15}$ cm$^3$molecule$^{-1}$).[38] The results of fitting also verify that the ozone pressures applied for the data in Figures 4 and 5 are small enough, so that this saturation effect should not appear in our kinetics measurements.

**Experimental uncertainty**

The error reported for each bimolecular reaction rate coefficient $k_2$ obtained from Equation (3) includes the errors of measuring the ozone pressure, [AH$_2$]$_0$ and the radius and fitting the gradient from the Raman intensity time profile. The percentage error of measuring the ozone pressure is fixed to 2.6%, which is the assumed maximum error of the measured UV absorption cross section of ozone at 250 nm estimated by literature.[48] The percentage error of measuring [AH$_2$]$_0$ is about 4%, which is attributed to fitting error of calibration curves and statistics of Raman intensities. The error of the droplet radius is assumed to be 0.15 μm, which is an averaged change of measured radii during the reaction progress, as the fitting error from the simulation (few nanometers) and statistics of data (~20 nm, see Figure 2) are significantly smaller. For the case of pH ≈ 2, the typical percentage error of fitted gradients is about 1%. Such small fitting error could probably be ascribed to the removal of CERS interferences and small Raman intensity fluctuations gained from concentrated AH$_2$ (~3.3 M on average) in aerosols. For the case of pH≈ 6, the typical percentage error of fitted gradients is about 24%. This relatively large fitting error may be attributed to large Raman intensity fluctuations due to the relatively low concentrations of AH$_2$ (~0.6 M on average) in aerosols and a potentially insufficient removal of CER interferences, which will be discussed later.

Overall, these errors were propagated into the determination of the error of $k_2$ by Equations S4 and S5 shown in ESI. As a result, the typical percentage error of the derived $k_2$ at pH $\approx$ 2 is about 12%, which is dominated by the uncertainties of reactant concentrations and radius. This value at pH $\approx$ 6 is about 36%, which is dominated by the fitting error of gradient. It should be noted that the errors of $k_2$ reported here do not include the systematic errors potentially associated with size changes of droplets. The time-varying droplet volume due to size change can alter the species concentrations and also the ionic strength during the progress of reaction, causing additional deviations to the value of $k_2$ determined by Equation (3), which assumes a fixed droplet volume. For the case of 3 μm droplet radius, the corresponding decrease of droplet volume due to radius drop of 0.15 μm is 14%. According to Equation (4), an increase of 14% in ionic strength can lead to an increase of about 32% or 7% in $k_2$ for pH $\approx$ 2 or 6, respectively, causing an overestimation of $k_2$. On the other hand, the enrichment of reactants, such as $AH_2$, during the reaction could prolong the reaction time and lead to an underestimation of $k_2$. Based on Equation (3), we could roughly estimate that such underestimation of $k_2$ is similar to the change of volume, such as 14%. As a result, the maximum systemic deviation of $k_2$ due to shrinking droplet volume is estimated to be about 18% overestimation or 7% underestimation for pH $\approx$ 2 or 6, respectively.

Finally, we would like to discuss the potential contributions of CERS interferences to the fitting error of gradient in Equation (3) and how to deal with them. As described before, we dealt with the CERS interferences *via* tracing the CERS peaks superposed on the molecular Raman band of interest and excluding the data points associated with such interference. Without such treatment, the fit in Figure 3 can significantly deviate from the time-dependent trend in intensity defined by Equation (3). However, we were not able to assign weak CERS signals which resemble the noise of spontaneous Raman spectra, and they could just manifest themselves as an intensity fluctuation in the Raman intensity time profile. This is probably one of main sources of the relatively large uncertainty of $k_2$ at pH ≈ 6. To further reduce such interference noise, we propose that it is actually desired to enhance CERS signals, so that they can be easily assigned and deconvoluted in data analysis. As the intensity of the CERS peak is proportional to the spontaneous Raman intensity at the same wavelength,[35] one feasible option is to enhance the spontaneous Raman signals *via* increasing the power or lowering the wavelength of the Raman excitation laser. In future works, we prefer to upgrade the single-beam AOT utilized in the present work to a dual-beam AOT with a counter-propagating geometry which has been demonstrated by several research groups,[49–51] as such dual-beam geometry can facilitate a stable trapping with a relatively large laser power.

Table 1. Reaction rate coefficient $k_2$ determined from the present work and previous measurements.

| Study | $I$ / M | free acid [a] / M$^{-1}$s$^{-1}$ | monoanion [b] / M$^{-1}$s$^{-1}$ |
|---|---|---|---|
| This work | 0.0 [c] | $(1.7 \pm 1.1) \times 10^5$ | $(1.1 \pm 0.6) \times 10^7$ |
| | 0.2 [d] | $(3.1 \pm 2.0) \times 10^5$ | $(1.2 \pm 0.6) \times 10^7$ |
| Giamalva et al.[6] | | $(6.9 \pm 2.3) \times 10^5$ | $(5.6 \pm 2.6) \times 10^7$ |
| Kanofsky and Sima[7] | ∼0.2 | $(5.6 \pm 0.4) \times 10^5$ | $(4.8 \pm 0.4) \times 10^7$ |
| Kermani et al.[9] | ∼0.2 | | $(5.5 \pm 0.4) \times 10^4$ |

*a*. pH = 1.9 for this work, and 2.0 for Giamalva et al. and Kanofsky and Sima.

*b*. pH = 6 for this work, 4.8 for Giamalva et al., 7.0 for Kanofsky and Sima and 7.4 for Kermani et al.

*c*. $k_0$ from the fitting results of Figures 4 and 5.

*d*. estimated *via* equation (4) and the fitted $k_0$ and $b'$.

**Comparing the results with literature**

Table 1 summarizes the bimolecular rate coefficients $k_2$ determined by this study and previous measurements. To compare with the literature values measured at ionic strength 0.2 M, we estimated the values of $k_2$ at this ionic strength from our results fitted with equation (4). For the case of free acid (pH < pKa), our result, $(3.1 \pm 2.0) \times 10^5$ M$^{-1}$s$^{-1}$, has the similar magnitudes as those of previous works, such as Giamalva et al.[6] and Kanofsky and Sima.[7] Such agreement with the literature values measured in bulk solution (Giamalva et al.[6]) implies that our observed reaction kinetics is dominated by diffusion-limited kinetics inside the droplets. For the case of monoanion (pH > pKa), our estimated value of $k_2$ at ionic strength 0.2 M, $(1.8 \pm 0.8) \times 10^7$ $M^{-1}s^{-1}$, has the same orders of magnitude with those measured by Kanofsky and Sima[7] or Giamalva et al.,[6] while our value is about three times less their values. Finally, our results exhibit the significant dependence of the reactivity with pH, agreeing with the results of these previous works. Such pH dependence can be attributed to the different reaction mechanisms of the $AH_2$ free acid (pH < pKa) and monoanion (pH > pKa) forms. On the other hand, the agreement with literature values also justifies the aerosol pH measurements of this study.

Table 2. Estimated diffuso-reactive lengths, $l$, for ozone under various conditions of different pH and concentrations of $AH_2$.

| Condition | $k_2$ / $M^{-1}s^{-1}$ | $[AH_2]$ / M | $l$ / nm |
|---|---|---|---|
| This work (pH ≈ 6) | $1.2 \times 10^7$ | 0.4-1.1 | 12-19 |
| This work (pH ≈ 2) | $3.1 \times 10^5$ | 1.6-5.6 | 32-60 |
| ELF (pH > pKa) | $1.2 \times 10^7$ | $4 \times 10^{-3}$ | ~190 |
| ELF (pH < pKa) | $3.1 \times 10^5$ | $4 \times 10^{-3}$ | ~1200 |

**Comparison with surface reaction kinetics**

The kinetics of the $AH_2$ + $O_3$ reaction at the air-water interface has been investigated, and the corresponding surface reaction rates were found to be at least two orders of magnitude larger than those from the bulk measurements.[1] To observe the surface kinetics in micro-droplets by means of AOT, this may ideally require the diffuso-reactive lengths to be within few nanometers.[26] Table 2 lists the predicted diffuso-reactive lengths with the experimental parameters used in this study. For the condition of this work at pH ≈ 2, the corresponding diffuso-reactive lengths are about few tens of nm, implying that any interface effect should not be observable in this study. Thus, the observations of the enhanced reactivities at pH ≈ 2 in this study could be mainly ascribed to the effect of ionic strength instead.

For the condition of this work at pH ≈ 6, the estimated diffuso-reactive lengths are reduced to about 12-19 nm, because of the significantly large value of bimolecular reaction rate coefficient $k_2$. However, even such small length may still be too large to observe the desired interface enhancement. Indeed, the maximum value of $k_2$ at pH ≈ 6 observed in this study is $(2.4 \pm 0.7) \times 10^8\,\mathrm{M^{-1}s^{-1}}$, which is only about three times larger than the literature value, but not the expected two orders of magnitude enhancement due to interface reaction. If we want to further reduce the diffuso-reactive length to few nanometers, we have to increase the concentration of $AH_2$ to over few molarities. On the other hand, the relatively high concentrations of sodium phosphate required for maintaining pH ≈ 6 also result in large ionic strengths in the aqueous $AH_2$ droplets. The effect of ionic strength to the reactive uptake of liquid phase aerosols have been investigated by several aerosol studies, such as the reactions of $SO_2$ with peroxides and the reactions of methoxyphenols with ozone.[52,53] As the data of $k_2$ can be reasonably modelled by equation (4), as shown in Figure 5, the observed enhancement of $k_2$ at pH ≈ 6 in this study is also ascribed to the effect of ionic strength.

The potential role of interfacial chemistry in aerosol reaction kinetics investigated here could also be examined by varying the droplet size to change the surface area-to-volume ratio. This approach has been often used in previous studies of aerosol kinetics to understand whether the heterogeneous reaction is dominated by volume- or surface-limited case, such as the studies on heterogeneous reactions of $N_2O_5$ on organic or inorganic aerosol particles.[54,55] These studies demonstrated that the volume-limited case can manifest the dependence of reactive uptake coefficients on particle size. In contrast, the surface-limited case does not exhibit such dependence.[54,55] For the present work, we plotted the bimolecular rate coefficients corrected to zero ionic strength ($ke^{-b'I}$, according to Equation (4)) at pH ≈ 2 and 6 as a function of droplet radius, as shown in Figures S6 and S7, respectively. These results exhibit no observable dependence of size, verifying that the observed kinetics of this work is dominated by the surface-limited case (diffuso-reactive length << radius).[54,55]

## Implications for ELF and bioaerosol

Several previous studies have suggested that the diffusions of potentially unreacted ozone or secondary oxidants generated from ozonolysis of antioxidants across the ELF could induce the oxidation damages and even the cell injury of biomembranes, while the type of secondary species, such as singlet $O_2$ or ascorbate ozonide, could depend on the acidity of ELF.[1,3] The findings of the present study indicate that the acidity can also affect the kinetics of the $AH_2$ oxidation reaction and the diffusion lengths of ozone. Table 2 summarizes the estimated diffuso-reactive lengths of ozone in ELF at two different pH, assuming that the ionic strength of ELF is 0.2 M.[56] Normally, the pH of ELF is about 6.9,[57] and the concentration of $AH_2$ is within few mM.[58] Thus, according to Table 2, the large reaction rage coefficient of the $AH_2$ + $O_3$ reaction at this pH can yield a relatively small diffuso-reactive length for ozone which could avoid the full penetration of ozone through the ELF and the direct contacts of ozone to the biomembranes. On the other hand, at higher acidities the significantly smaller reaction rate coefficients of the $AH_2$ + $O_3$ reaction can yield larger diffuso-reactive lengths, which can be even much larger than the depth of ELF. As a result, the higher acidity of ELF provided by various pathologies or inhaled particular matters can not only generate more AOZ,[1] but also result in the deeper penetrations of AOZ and ozone through the ELF and thus the higher chances of their direct contacts to the aerial biosurfaces.

The aqueous $AH_2$ aerosols studied in the present work can also be regarded as prototype bioaerosols. Humans can exhale small droplets of airway-lining fluids, such as ELF, during normal breathing, coughing or sneezing. These exhaled bioaerosols may carry airborne pathogens, and a detailed understanding about the influences of aerosol chemistry and microphysics to these pathogens in aerosols becomes crucial recently.[59–61] The airway-lining fluids typically contain trace amounts of antioxidants, such as $AH_2$, uric acid and reduced Glutathione, besides salts and proteins. The findings of this study suggest that the concentrations of above species and the ionic strength inside the bioaerosol droplets could increase at least about a few times due to evaporation, while the bioaerosols may still remain the similar pH at 6.9. As a result, the concentrated

AH$_2$ in bioaerosol remains the high reactivities and yields small diffuso-reactive lengths for ozone (~120 nm, assuming enriched [AH$_2$] = 10 mM), compared to the typical size of bioaerosol (< 1-10 μm). Thus, the antioxidants, such as AH$_2$, actually become to protect the pathogens inside bioaerosol against ozone from environments for a period of time. The reaction time required to scavenge enriched AH$_2$ (~10 mM) in bioaerosols (~5 μm in size) with indoor ozone (~10 ppb) could take about a few hours. For the more "toxic" atmospheric ozone (~50 ppb), the scavenge time is then reduced to about a half hour.

Finally, we would like to justify the usage of the relatively high pressure of ozone in this work (~2 ppm on average), when comparing to ambient ozone (~50 ppb). Firstly, the experimental results of pressure dependence in this work justify that the typical ozone pressure used here is sufficiently low, so that the corresponding value of $k_2$, 3.0 × 10$^6$ M$^{-1}$s$^{-1}$ (at pH ≈ 2, ozone pressure = 1.4 ppm and ionic strength = 1 M, see Figure 6), is already very close to that at the low pressure limit, 3.1 × 10$^6$ M$^{-1}$s$^{-1}$. This means that the ozonolysis of AH$_2$ droplets at both ~2 ppm and ~50 ppb ozone pressures can be approximated to the same special case of heterogeneous reaction kinetics, i.e., diffusion-limited case characterized by Equation (3) and diffuso-reactive length, without the need to include any ozone-pressure dependent kinetics, such as Langmuir-Hinshelwood-type mechanism. As a result, the main features of heterogeneous reaction kinetics investigated by this work should be the same as those for ambient ozone. On the other hand, the primary purpose of higher ozone pressures used in this laboratory investigation is to shorten the entire reaction time, allowing for building statistics via repeating more measurements on reasonable experimental time scales.

## Conclusions

In this work, we investigated the kinetics of aqueous AH$_2$ reaction in micron-sized droplets with gaseous ozone by means of aerosol optical tweezers, at different ozone pressures, ionic strengths and pH. Particularly, this study demonstrates that the kinetics and pH of single aerosol droplets can be determined simultaneously. The measured bimolecular reaction rate coefficients at low ozone pressures and low ionic strengths agree with those from previous bulk measurements, indicating that the observed aerosol reaction kinetics can be solely explained in terms of liquid phase diffusion and $AH_2$ + ozone reaction, and no necessary to include any surface effect. We found that the measured bimolecular reaction rate coefficients exhibit a Langmuir-Hinshelwood dependence on ozone pressures, ascribed to the heterogeneous nature of such aerosol reaction. We also found that the measured bimolecular reaction rate coefficients have positive correlations with ion strengths. This study also confirms that the reaction rate coefficients of the $AH_2$ + ozone reaction at pH < pKa, such as pH $\approx$ 2, can be about two orders of magnitude smaller than that at pH > pKa, such as pH $\approx$ 6, agreeing with previous bulk measurements. These results imply that the higher acidity in airway-lining fluids could cause the lower reactivity of $AH_2$ to ozone. On the other hand, $AH_2$ in exhaled bioaerosols could actually protect pathogens within against the oxidative damages caused by atmospheric ozone for a period of time, such as a few hours.

## Conflicts of interest

There are no conflicts of interest to declare.

## Acknowledgements

This work was supported by the Ministry of Science and Technology, Taiwan (MOST107-2113-M-110-004-MY3 and MOST109-2113-M-110-010-) and NSYSU-KMU joint research project (NSYSUKMU 108-I002). We also thank for the financial supports from Aerosol Science Research Center, NSYSU, Taiwan. Finally, we thank Prof. Adam J. Trevitt for the tutorial of manipulating aerosols, and we also thank Prof. Chia C. Wang and referees of this manuscript for their very helpful comments.

## Notes and references